\documentclass[preprint,aps,epsfig]{revtex}
\begin{document}

\title{ Density profiles of supernova matter and determination of neutrino parameters}

\author{Shao-Hsuan Chiu\thanks{schiu@mail.cgu.edu.tw}}
\address{Physics Group, C.G.E., Chang Gung University, 
Kwei-Shan 333, Taiwan}

\maketitle

\newif\iftightenlines\tightenlinesfalse
\tightenlines\tightenlinestrue

\begin{abstract}

The flavor conversion of supernova neutrinos
can lead to observable signatures related to the unknown neutrino parameters. 
As one of the determinants in dictating the efficiency of resonant flavor conversion,
the local density profile near the MSW resonance in a supernova environment is, 
however, not so well understood.
In this analysis, variable power-law functions are adopted to 
represent the independent local density profiles near the locations of resonance.
It is shown that the uncertain matter density profile in a supernova, 
the possible neutrino mass hierarchies, and the undetermined 1-3 mixing 
angle would result in six distinct scenarios in terms of the survival
probabilities of $\nu_{e}$ and $\bar{\nu_{e}}$.
The feasibility of probing the undetermined neutrino mass
hierarchy and the 1-3 mixing angle
with the supernova neutrinos is then examined using 
several proposed experimental observables.
Given the incomplete knowledge of the supernova matter profile,
the analysis is further expanded to incorporate
the Earth matter effect. 
The possible impact due to the choice of models, 
which differ in the average
energy and in the luminosity of neutrinos, 
is also addressed in the analysis.

\end{abstract}


\pagenumbering{arabic}

%%%%%%%%%%%%%%%%%%%%%%%%%%%%%%%%%%%%%%%%%%%%%%%%%%%%%%%%%%%%%%%%%%%%%%%%
%%%%%%%%%%%%%%%%%%%%%%%%%%%% MAIN TEXT %%%%%%%%%%%%%%%%%%%%%%%%%%%%%%%%%

%%%%%%%%%%%%%%%%%%%%%%%%%% SECTION 1 %%%%%%%%%%%%%%%%%%%%%%%%%%%%%%%%%%%%%%%%

\section{Introduction}

Definite evidences from recent 
experiments~\cite{sk:99,sno:01,sk:2004,kam:03,k2k:03,chooz:99,palo:00} 
have confirmed that the neutrinos are massive and mixed.
Part of the intrinsic neutrino properties, 
such as the neutrino mass hierarchy and the 
1-3 mixing angle, however, remain undetermined.
The need for a better knowledge of the neutrino physics
has motivated plenty of phenomenological analyses
on whether and how the unknown neutrino properties can be reliably probed 
with the next generation experiments involving neutrinos from both 
the astrophysical and the terrestrial 
sources~\cite{dighe:00,luna:03,barger:02,dighe:03,dav:05,fogli:02,ska:07}.

The core-collapse supernovas represent a unique
type of neutrino source 
in that they emit neutrinos of all three flavors 
with a characteristic energy range in a time scale distinct from
that of the neutrinos emitted from the sun, the atmosphere, and the terrestrial sources.
The rich physical content involved in the 
environment further makes the supernova an important platform 
for the study of neutrino properties. 
In particular, the neutrino burst from a supernova has long been considered as a  
promising tool for probing the undetermined 
neutrino parameters, despite the fact that the complexity arising from the 
unavoidable astrophysical uncertainties 
may result in ambiguity in the interpretation of the observed events.

The standard MSW formulation~\cite{w,ms} suggests that 
as the supernova neutrinos propagate outward, the variation of  
density profile can have an impact on the resonant flavor conversion. 
One of the difficulties in probing the unknown neutrino properties
with the supernova neutrinos arises from lack of knowledge to the 
matter density profile in a supernova. 
The inverse power-law density, $\rho \sim r^{-3}$,
of the progenitor star is usually
adopted in the literature as the density distribution through which the supernova 
neutrinos propagate outwards.
However, whether the global density distribution $\rho \sim r^{n}$ with 
a fixed power can provide a justifiable connection
between the dynamics of resonant flavor conversion and the expected
neutrino events at the detectors remains uncertain. 
In addition, one may easily question the validity of this over-simplified model 
if, among others, the time-dependent matter distribution
driven by the shock waves and the associated effects 
(see, $e.g.$, Ref.~\cite{shock,shock-1,shock-2,shock-3,barger:05,chr:06}) 
are considered.
It is shown~\cite{chiu} that near the MSW resonance, even a relatively small deviation 
of the power from $n=-3$ could have a sizable
impact on the interpretation of the neutrino parameters.
Furthermore, no evidence suggests that the
density distributions near both the higher resonance and the lower resonance
should be described by a similar density profile.

Another major source of astrophysical uncertainty 
in analyzing the supernova neutrinos would be the spectral parameters. 
Distinct predictions to the average
energy and the luminosity of neutrinos exist among models.  
Thus, the possible impact due to the uncertain spectral parameters
also needs to be addressed in an analysis. 

Independent and variable power-law functions 
are assumed in this work as the matter profiles near the locations of MSW resonance. 
In addition, physical observables derived
from the expected neutrino events at a water Cherenkov detector are proposed.
It is the aim of this work to analyze the outcomes due to 
the uncertain density profile and the choice of spectral models,
and to investigate whether and how the experimental observables 
would provide hints for the determination of the unknown neutrino parameters. 
The analysis is also extended to include the Earth matter effect. 
This work is organized as follows.  Section II summarizes the 
general features of the neutrino fluxes emitted by a core-collapse supernova
and that of the relevant detection processes at a water Cherenkov detector. 
In section III,
the variable power-law function is proposed to account
for the possible uncertainty in the matter profile near the  
MSW resonance. In addition, observables are constructed and examined.
Focus has been placed on whether and how these observables would
resolve the various scenarios arising from the 
uncertainty in the matter profile, the possible neutrino mass hierarchies, 
and the undetermined mixing angle $\theta_{13}$.
In Section IV, the general formulation is expanded to incorporate
the regeneration effect due to the Earth matter.  In Section V,
the observables proposed in Section III are re-analyzed
with the Earth matter effect
included. Section VI is devoted to analyzing the impact 
due to the choice of spectral models.  We then summarized this work in Section VI.

%%%%%%%%%%%%%%%%%%%%%%%% section II %%%%%%%%%%%%%%%%%%%%%%%%%%%%%%%%%%%%%%%%%
\section{Supernova Neutrinos at the detector}
%%%%%%%%%%%%%%%%%%%%%%%%%%%%%%%%%%%%%%%%%%%%%%%%%%%%%%%%%%%%%%%%%%%  
 
A core-collapse supernova emits neutrinos of all three flavors
on a typical time scale of roughly ten seconds.
Within a factor of two, the total luminosity 
is approximately equi-partitioned into each flavor. The mean energy of each 
neutrino flavor is determined by
its interacting strength with matter. 
It is believed that the neutrino spectra are not exactly thermal; they
are usually modeled by the pinched Fermi-Dirac distributions.
The flux of flavor $\nu_{\alpha}$ reaching the Earth surface can be written as
%%%%%%%%%%%%%%%%%%%%%%%%%%%%%%
\begin{equation}\label{eq:f0}      
 F^{0}_{\alpha}=\frac{L_{\alpha}}{4\pi r^{2} T^{4}_{\alpha} F_{3}(\eta_{\alpha})}
 \frac{E^{2}_{\alpha}}{e^{[(E_{\nu}/T_{\alpha})-\eta_{\alpha}]}+1},
\end{equation} 
%%%%%%%%%%%%%%%%%%%%%%%%%%%%%%%%
where $L_{\alpha}$ is the luminosity of the neutrino flavor $\nu_{\alpha}$, 
$r$ is the distance to a supernova, $T_{\alpha}$ is the effective 
temperature of $\nu_{\alpha}$ inside the respective neutrinosphere, $E_{\alpha}$ is
the energy of $\nu_{\alpha}$, and $\eta_{\alpha}$ is the pinching parameter.
The normalization factor $F_{3}(\eta_{\alpha})$ is given by
%%%%%%%%%%%%%%%%%%%%%%%%%%%%%%%%%%%
\begin{equation}\label{eq:f3}
F_{3}(\eta_{\alpha}) \equiv -6 Li_{4}(-e^{\eta_{\alpha}}),
\end{equation}
%%%%%%%%%%%%%%%%%%%%%%%%%%%%%%%%%%%%%%%
with the polylogarithm function $Li_{n}(z)$ defined as
%%%%%%%%%%%%%%%%%%%%%%%%%%%%%%%%%%%%%%
\begin{equation}\label{eq:Li}
Li_{n}(z) \equiv \sum_{k=1}^{\infty} \frac{z^{k}}{k^{n}}.
\end{equation} 
%%%%%%%%%%%%%%%%%%%%%%%%%%%%%%%%%%%%%%%%
Note that the temperature $T_{\alpha}$ is related to 
the mean energy $\langle E_{\alpha} \rangle$ by
%%%%%%%%%%%%%%%%%%%%%%%%%%%%%%%%%%
\begin{equation}\label{eq:ea}
\langle E_{\alpha} \rangle=3 \frac{Li_{4}(-e^{\eta_{\alpha}})}
{Li_{3}(-e^{\eta_{\alpha}})}T_{\alpha}.
\end{equation}
%%%%%%%%%%%%%%%%%%%%%%%%%%%%%%%%%%%%%%%%

%%%%%%%%%%%%%%%%%%%%%%%%%%%%%%%%%%%%%%%
%revised
%%%%%%%%%%%%%%%%%%%%%%%%%%%%%%%%%%%%%%%%%%%%%%%%%%
                                                 % 
As an illustration, the spectral parameters published by the
Lawrence Livermore (LL) group~\cite{LL} will be adopted in the analysis:
$\langle E_{\nu_{e}} \rangle : \langle E_{\bar{\nu}_{e}} \rangle :
\langle E_{\nu_{x}} \rangle \approx 4:5:6$ with 
$\langle E_{\bar{\nu}_{e}} \rangle =15$ MeV, and
$L_{\nu_{e}}/L_{\nu_{x}}$ = 2.0, $L_{\bar{\nu}_{e}}/L_{\nu_{x}}=1.6$,
where $\nu_{x}=\nu_{\mu}$, $\nu_{\tau}$, $\bar{\nu}_{\mu}$, and $\bar{\nu}_{\tau}$.
The pinching parameters
$\eta_{\nu_{e}} = \eta_{\bar{\nu}_{e}} = \eta_{\nu_{x}}=3.0$ are assumed.
As for the neutrino parameters, we choose  
$\delta m^{2}_{21}=7.0 \times 10^{-5}$ eV$^{2}$, 
$|\delta m^{2}_{31}|=3.0 \times 10^{-3}$ eV$^{2}$, and
$\sin^{2}2\theta_{12}=0.81$.
Note that the ongoing work by the Garching group~\cite{Ga,Gb}
predicts distinct neutrino fluxes and average energies from that
of the LL group.  The comparison will be performed and
discussed in Section VI.

In a water Cherenkov detector, such as the 
Super-Kamiokonde (see, $e.g.$, ~\cite{sk:99}),
the upgraded Hyper-Kamiokonde~\cite{hk}, 
and other facilities that are under planning or construction~\cite{detectors},
the event rate induced by the isotropical inverse $\beta$-decay process 
\begin{equation}\label{eq:beta}
\bar{\nu}_{e}+p \rightarrow n+e^{+}
\end{equation}
dominates over that of other channels.  
In this analysis, the isotropical CC processes:
\begin{equation}\label{eq:F}
\nu_{e}+O \rightarrow F+e^{-}, 
\end{equation}

\begin{equation}\label{eq:N}
\bar{\nu}_{e}+O \rightarrow N+e^{+}, 
\end{equation} 
and the highly directional events induced by 
$\nu_{\alpha}+e^{-}$ ($\alpha=e,\mu,\tau$) and
$\bar{\nu}_{\alpha}+e^{-}$ scattering
will also be included.

%%%%%%%%%%%%%%%%% section III %%%%%%%%%%%%%%%%%%%%%%%%%%%%%%%%%%%%%%%%%%%%%%%%%
\section{Constructing observables from the neutrino signals}

%%%%%%%%%%%%%%%%%%%%%%%%%%%%%%%%%%%%%%%%%%%%%%%%

\subsection{The formulation}

The neutrinos would likely go through multiple flavor transitions
in a supernova at distinct density scales.
A single power-law function,
$\rho \sim r^{-3}$ or $\rho \sim r^{n}$ in general, is usually adopted
in the literature as the density profile of the supernova matter.
When the neutrinos travel through
a complex environment such as the matter of a supernova, however, 
it would seem more plausible in an analysis 
to allow independent and variable local density profiles 
near the resonant locations.  

If the matter distribution is modeled as
layers of matter along the path of neutrinos in a specific direction, 
a simple power-law function
can be adopted as the local profile within a layer: 
\begin{equation}
\rho (r)_{k} =c_{k} r^{n_{k}}, 
\end{equation}
where $k$ stands for the $k$-th layer,
$r^{n_{k}}$ describes the matter profile within the $k$-th layer,
and $c_{k}$ denotes the magnitude of the component $r^{n_{k}}$. 
The density profile near the location of resonance can then be written as
\begin{equation}
\rho (r)_{k_{0}} = c_{k_{0}} r^{{n}_{k_{0}}},
\end{equation}
where the $k_{0}$-th layer stands for the resonance layer of the flavor conversion.
Note that in this approach the power can vary
from layer to layer to account for the uncertainty of profile originating
from various types of sources. The thickness of the layers,
which do not take part in the formulation of this analysis, 
can be uneven and unrestricted,
with the assumption that a specific power $n$ properly describes the density profile 
within a specific layer.
Furthermore, the variable power and thickness also allow the 
deviation of the matter density from spherical symmetry
and the inhomogeneity of matter
distribution along different directions.

One may also employ functions of more than one polynomial term.
The extra polynomial terms would provide more details about the density
profile in a global scale.  However, the trade-off is that
they also introduce more undetermined coefficients for the extra terms.
Since one only concerns the local density profile near the MSW layer, 
the detailed shape and variation
of the density in a global scale would be unimportant, and the extra terms 
with their undetermined coefficients would become redundant in the analysis.
Thus, the power-law function with an adjustable $n$ 
would be suitable and much simpler
for the analysis on how the variation of local density profile, 
rather that the global profile, would impact the MSW resonance.

In the following, one may simply focus on the density shapes of matter
in the resonance layers:  
$\rho (r) = c_{l} r^{n_{l}}$ for the lower resonance,  
and $\rho (r) = c_{h} r^{n_{h}}$ for the higher resonance.
The variation of $n_{l}$ and $n_{h}$ would be expected to alter the 
efficiency of neutrino flavor transition. 

Along this approach, the crossing probabilities at the resonance can be modified as
\begin{equation}\label{eq:pc} 
 P_{k} = \frac{\exp(-\frac{\pi}{2} \gamma_{k} F_{k})-
 \exp(-\frac{\pi}{2} \gamma_{k} \frac{F_{k}}{\sin^{2} \theta_{ij}})}
 {1-\exp(-\frac{\pi}{2} \gamma_{k} \frac{F_{k}}{\sin^{2} \theta_{ij}})},
   \end{equation}
where $k=h$ or $l$, $\theta_{ij}$ is the mixing angle, 
$F_{k}$ is the correction factor to a non-linear profile,
and the adiabaticity parameter $\gamma_{k}$ takes the form
\begin{equation}\label{eq:gamma}
   \gamma_{k}=\frac{1}{2|n_{k}|}(\frac{\delta m^2_{ij}}{E_{\nu}})^{1+\frac{1}{n_{k}}}
   (\frac{\sin^2 2\theta_{ij}}{\cos 2\theta_{ij}})
   (\frac{\cos 2\theta_{ij}}
   {2\sqrt{2} G_{F}\frac{Y_{e}}{m_{n}} c_{k}})^{\frac{1}{n_{k}}},
\end{equation}
where $\delta m^2_{ij} \equiv |m^{2}_{i}-m^{2}_{j}|$, 
$G_{F}$ is the Fermi constant, $E_{\nu}$ is the neutrino energy, 
$Y_{e}$ is the electron number per baryon, $m_{n}$ is the baryon mass,
and the scale $c_{k}$ varies very weakly with $r$ over the range 
$10^{12} g/cm^{3} < \rho < 10^{-5} g/cm^{3}$.
Note that $\bar{P}_{h}=P_{h}$, and that $\bar{P}_{l}$ 
can be obtained directly from $P_{l}$ by replacing $\theta_{ij}$ with 
$\pi/2-\theta_{ij}$.

The probabilities that $\nu_{e}$ and $\bar{\nu}_{e}$ 
survive the flavor conversion are given respectively by
\begin{equation}\label{eq:pe}
       P_{nor} \simeq U_{e1}^{2} P_{l} P_{h} + U_{e2}^{2} (1-P_{l}) P_{h} +
                 U_{e3}^{2} (1-P_{h}),
\end{equation} 
   
\begin{equation}\label{eq:pa}
   \bar{P}_{nor} \simeq U_{e1}^{2} (1-\bar{P}_{l}) + U_{e2}^{2} \bar{P}_{l},
\end{equation} 
for the normal hierarchy, and

\begin{equation}\label{eq:pei}
    P_{inv} \simeq U_{e2}^{2} (1-P_{l}) + U_{e1}^{2} P_{l},
\end{equation}
   
\begin{equation}\label{eq:pai}
    \bar{P}_{inv} \simeq U_{e2}^{2} \bar{P}_{l} \bar{P}_{h}+
                  U_{e1}^{2} (1-\bar{P}_{l}) \bar{P}_{h} + 
                 U_{e3}^{2} (1-\bar{P}_{h}),
\end{equation}
for the inverted hierarchy.
Eqs.~(\ref{eq:pc}) and ~(\ref{eq:gamma}) imply that
the variation of neutrino energy plays an insignificant
role~\cite{chiu}, as compared to the variations of $n_{k}$ and the 1-3 mixing angle,
in affecting the survival probabilities.
In addition, the two extreme values of $P_{h}$: 
$P_{h} \sim 1$ (non-adiabatic) and $P_{h} \sim 0$ (adiabatic), 
are given by the conditions: $P_{h} \sim 1$ if $g <1$,
and $P_{h} \sim 0$ if $g >1$, where the function $g$ relates $n_{h}$
and $\theta_{13}$:
\begin{equation}\label{eq:g}
  g=g(n_{h},\theta_{13}) \simeq \frac{2.3 \times 10^{-4}}{|n_{h}|}
  (37.6 \times 10^{-30} \times \cos 2\theta_{13})^{\frac{1}{n_{h}}} 
  \frac{\sin^{2} 2\theta_{13}}{\cos 2\theta_{13}}.
\end{equation} 
Furthermore, Eqs.~(\ref{eq:pc}) and ~(\ref{eq:gamma}) also suggest that 
the arbitrary $n_{l}$ gives rise to only two distinct
values of $P_{l}$ ($\bar{P}_{l}$): $P_{l} \simeq 0$ ($\bar{P}_{l} \simeq 0$)
for $n_{l}>-5$, and $P_{l} \simeq \cos^{2}\theta_{12}$ 
($\bar{P}_{l} \simeq \sin^{2}\theta_{12}$) for $n_{l}<-5$.
 
As summarized in Table 1, 
the variation of $n_{k}$ ($k=l, h$), the undetermined $\theta_{13}$, and
the possible mass hierarchies would lead to six distinct scenarios 
in terms of the survival probabilities.
In the first row of Table 1, the quantities in square brackets
denote the variables, $i.e.$, $P_{l}$ and $\bar{P}_{l}$ are determined by
the variable $n_{l}$; $P_{h}$ is determined by $n_{h}$ and $g$;
and the total survival probabilities, 
$P$ and $\bar{P}$, are determined by all $n_{l}$, $n_{h}$, $g$,
and the mass hierarchy.  
The last column labels the six different scenarios
in terms of the possible combinations formed by $P$ and $\bar{P}$.
Note that the variation of $n_{k}$ would 
impact the probabilities differently according to whether $n_{k}>-5$ or $n_{k}<-5$.
Furthermore, the outcomes due to $\rho \sim r^{-3}$, which is usually assumed in
the literature, correspond
to those due to ($n_{l}>-5$, $n_{h}>-5$) in Table 1.

Some qualitative observations are outlined as follows. 
(i) No information on the mass hierarchy or on $\theta_{13}$ is available
if either scenario I or scenario IV is observed.  
(ii) If either II or III is observed, 
then the lower bound on $\theta_{13}$ would be available.
(iii) Likewise, if either V or VI is observed, 
then the lower bound on $\theta_{13}$ would be available.

The neutrino events may be categorized into
isotropical ( Eqs.~(\ref{eq:beta}), ~(\ref{eq:F}), ~(\ref{eq:N})) 
and directional (neutrino scattering) events at the water Cherenkov detector.
The expected time-integrated number spectra for both the 
isotropical and the directional events are shown in Figs. 1(a) and 1(b), 
respectively.  It can be seen from Fig. 1(a) that 
three groups of nearly degenerate spectra would form for the isotropical
events: (I,V), (II,IV),
and (III,VI). Thus, the isotropical events alone fail 
to provide all the needed clues for separating the six scenarios. 
On the other hand, while the directional event
rates shown in Fig. 1(b) amount to only a tiny fraction of the isotropical ones, 
they could provide the rest of the clues for
lifting the above degeneracy.
As a result, combining both the isotropical and the directional events 
is inevitable in the construction of feasible observables.

The $\nu_{e}$ and $\bar{\nu}_{e}$ fluxes arriving at the surface of Earth,
$F_{e}$ and $F_{\bar{e}}$, 
are related to the original ones, $F^{0}_{e}$ and $F^{0}_{\bar{e}}$,
through the survival probabilities $P$ and $\bar{P}$
for $\nu_{e}$ and $\bar{\nu}_{e}$, respectively:
\begin{equation}\label{eq:f}
F_{e}=F^{0}_{e}+(1-P)(F^{0}_{x}-F^{0}_{e}),
\end{equation}
\begin{equation}\label{eq:fa}
F_{\bar{e}}=F^{0}_{\bar{e}}+(1-\bar{P})(F^{0}_{\bar{x}}-F^{0}_{\bar{e}}).
\end{equation}
The expected total event number $N_{\alpha}$ induced by the flavor $\nu_{\alpha}$ 
at the detector is given by
\begin{equation}\label{eq:event}
N_{\alpha}=\frac{N_{t}}{4\pi r^{2}} \int \sum_{i} [F_{\alpha} \sigma_{\alpha i}] dE,
\end{equation}
where $N_{t}$ is the number of targets at the detector, $r$ is the distance to
the supernova, and $\sigma_{\alpha i}=\sigma_{\alpha i}(E)$ is the cross section for
$\nu_{\alpha}$ in a particular reaction channel $i$. 
The detection efficiency is assumed to be one.
Note that the regeneration effect due to the Earth matter will be expected to alter
the survival probabilities, and thus
the observed event rates.
We shall first concentrate on building the observables without considering 
the Earth regeneration effect.  A more detailed analysis incorporating
the Earth effect will follow.

%%%%%%%%%%%%%%%%%%%%%%%%%%%%%%%%%%%%%%%%%%%%%%%%%%%%%%%%%%%
%%%%%%%%%%%%%%%%%%%%%%%%%%%%%%%%
\subsection{Observables derived from neutrino signals}

Statistically, the predominant isotropical channels can provide
the most satisfying events for the analysis.
Fig. 1(a) suggests that one may obtain the
optimized results by examining $R_{1}$, which is defined as the ratio
of the event rates in the medium and low energy range ($E<40$ MeV) 
to that in the high energy tail ($E>40$ MeV):

\begin{equation}\label{eq:R1}
R_{1} \equiv \frac{N_{iso}(E<40 MeV)}{N_{iso}(E>40 MeV)}.
\end{equation}
This ratio can be reduced to
\begin{equation}\label{eq:R1r}
R_{1} \simeq 2.62\frac{1+0.30\bar{P}-0.008P}{1-0.88\bar{P}-0.028P}.
\end{equation}
This expression leads to the following possible values for $R^{J}_{1}$, where $J$
labels the six scenarios:
(i) $R_{1}^{I}=R_{1}^{V} \sim 6.6$.
(ii) $R_{1}^{II}=R_{1}^{IV} \sim 8.8$, 
(iii) $R_{1}^{III}=R_{1}^{VI} \sim 2.6$.
It is seen that the values of $R_{1}$ for the group (III,VI) are smaller
than the others by a factor of $2 \sim 3$ and can be easily singled out. 

From a different point of view, one notes that
the inverted mass hierarchy implies the following possible results:
(i) $R_{1} \sim 6.6$, which predicts $n_{l}<-5$, $n_{h}<-5$,
while the information of $\theta_{13}$ is unavailable. 
(ii) $R_{1} \sim 8.8$, which predicts $n_{l}>-5$, $n_{h}>-5$, $g<1$,
and thus the upper bound $\sin^{2}\theta_{13} \sim 10^{-2}$.  
(iii) $R_{1} \sim 2.6$, which predicts $n_{l}>-5$, $g>1$, 
and thus the lower bound $\sin^{2}\theta_{13} \sim 10^{-2}$. 
On the other hand, 
the following would be possible if the mass hierarchy is normal:
(i) $R_{1} \sim 8.8$, which predicts $n_{l}>-5$, 
while $n_{h}$ and $\theta_{13}$ are undetermined.
(ii) $R_{1} \sim 2.6$, which only predicts $n_{l}<-5$.

To establish the $P$-sensitive observables 
that resolve the degenerate scenarios in the same group,
we now examine the $\nu_{e}$-dominating directional events.
To optimize the results, one may define the ratio of the directional events for 
$E<25$ MeV to that for $E>25$ MeV as $R_{2}$:
\begin{equation}\label{eq:R2}
R_{2}\equiv \frac{N_{dir}(E<25 MeV)}{N_{dir}(E>25 MeV)}.
\end{equation}
This ratio can be reduced to 
\begin{equation}\label{r2}
R_{2} \simeq 0.98 \frac{1+0.23 \bar{P} +1.19 P}{1-0.086 \bar{P} -0.36 P},
\end{equation}
which leads to the following:
$R_{2}^{I} \sim 2.46$, $R_{2}^{II} \sim 1.22$, $R_{2}^{III} \sim 1.45$, 
$R_{2}^{IV} \sim 1.76$, $R_{2}^{V} \sim 1.17$,
and $R_{2}^{IV} \sim 2.13$.

It is seen that the six scenarios lead to distinct observable results in terms of the
combination ($R_{1}$, $R_{2}$).  The observable $R_{1}$ would reduce the six
scenarios into three degenerate groups: (I,V), (II,IV), and (III,VI). 
Observing $R_{2}$ would further remove the degeneracy between the scenarios
in each group.
Thus, each scenario can be characterized by a unique
set of observables $(R_{1},R_{2})$, which should act as a feasible reference in
removing the degeneracy among the six scenarios.  

%%%%%%%%%%%%%%%%%%%%%%%%%%%%%%%%%%%%%%%%%%%%%%%%%%%%

In searching for useful observables other than
$R_{1}$ and $R_{2}$, one notes from Fig. 1 that 
in terms of the height and the width, 
the possible spectra exhibit the most
pronounced differences in the region near the peaks.  
The ratio formed by the peak value of a specific spectrum ($F_{max}$)
and the width at half peak value ($\Gamma_{1/2}$) deserves some attention.
One may introduce the double ratio involving both
the isotropical and the directional events:
\begin{equation}
R_{3} \equiv \frac{(F_{max}/\Gamma_{1/2})_{iso}}{(F_{max}/\Gamma_{1/2})_{dir}}.
\end{equation}
The six scenarios lead to the following possible values: 
$R_{3}^{I} \sim 16.7$,  $R_{3}^{II} \sim 49.5$, $R_{3}^{III} \sim 22.6$,
$R_{3}^{IV} \sim 35.5$, $R_{3}^{V} \sim 40.4$, and $R_{3}^{VI} \sim 11.9$  

Furthermore, one may define another observable $R_{4}$
involving both the isotropical the directional events in the medium 
energy range, $13 MeV < E < 27 MeV$:
\begin{equation}
R_{4} \equiv \frac{N_{ios}(13 MeV<E<27 MeV)}{N_{dir}(13 MeV<E<27 MeV)},
\end{equation}  
which leads to
\begin{equation}
R_{4} \simeq 19.86 \frac{1+1.05 \bar{P}+0.0013 P}{1+0.16 \bar{P} +0.59 P},
\end{equation}
and $R_{4}^{I} \sim 22.3$,  $R_{4}^{II} \sim 31.3$, $R_{4}^{III} \sim 17.0$,
$R_{4}^{IV} \sim 27.2$, $R_{4}^{V} \sim 29.5$, and $R_{4}^{VI} \sim 14.7$

If the neutrino energy spectrum is reconstructed, one may 
examine yet another observable $R_{5}$, which involves the average
energies of both the isotropical and the directional spectra:
$R_{5} \equiv \langle E \rangle_{iso}/\langle E \rangle_{dir}$.
The combined average energy $\langle E \rangle_{i}$ is defined as
\begin{equation}\label{eq:ae}
\langle E \rangle_{i} \equiv \frac{\sum_{\alpha}[\int E (\frac{dN_{\alpha i}}{dE}) dE]}
{\sum_{\alpha} N_{\alpha i}},
\end{equation}
where $i$ stands for isotropical or directional, 
$\frac{dN_{\alpha i}}{dE}$ is the time-integrated number spectrum,
$N_{\alpha i}$ is the event number,
and the summation over $\alpha$ 
includes all the possible isotropical or directional 
channels induced by $\nu_{\alpha}$.
Further reduction of $R_{5}$ yields
\begin{equation}\label{eq:R5r}
R_{5} \simeq 1.26 \frac{(1-0.37\bar{P}-0.018 P)(1+0.71 \bar{P}+0.46 P)}
{(1-0.023 \bar{P}-0.014 P)(1-0.002\bar{P}-0.043 P)},
\end{equation}
and the values of $R_{5}$ for the six scenarios are given by:
$R_{5}^{I} \sim 1.27$, $R_{5}^{II} \sim 0.99$, $R_{5}^{III} \sim 1.41$,
$R_{5}^{IV} \sim 1.10$, $R_{5}^{V} \sim 1.04$, and $R_{5}^{VI} \sim 1.56$.  

The uncertainty arising from the density profile can leave observable
signatures on the neutrino events through the resonant 
flavor transition. 
For a given set of input spectral parameters, 
the proposed experimental observables may be combined to form  
useful references for removing the degeneracy among scenarios.
However, it should be noted that whether the unknown neutrino parameters
could be definitely determined remains as an open question as other 
possible sources of uncertainty may blur or even wash out the signatures.  
We shall discuss this issue in sections VI and VII.

%%%%%%%%%%%%%%%%%%%%%%%%%%%%%%%%%%%%%%%%%%%%%%%%%%%%%%%%%%%%%%%%%%
%%%%%%%%%%%%%%%%%%%%%%%%%%%%%%%%%%%%%%%%%%%%%%%%%%%%%%%%%%%%%
\section{Including the Earth matter effect}

The neutrinos emitted from a supernova arrive at the
Earth surface in mass eigenstates.  
Since the matter density of Earth
is approximately the same order of magnitude as that near
the lower resonance in a typical supernova,
the neutrinos could oscillate again 
if they cross Earth prior to the detection.
This regeneration effect in the Earth matter
would be expected to alter the neutrino fluxes arriving at the detector.
The observable effects due to this regeneration process have been 
intensively studied~\cite{ls:01}.
It is realized that in general the Earth effects are signaled by 
an oscillatory modulation on 
the neutrino spectrum that is expected when the regeneration effect is missing.
Given the uncertainty in matter density profile
of a supernova, we intend here to further 
examine the feasibility of the proposed observables under the influences of
the Earth regeneration effect.

The regeneration effect depends on several factors: 
the neutrino parameters,
the neutrino energy, the density shape of Earth matter, 
and the location of the detectors (or, the incident direction of the neutrinos).
The $\nu_{e}$ and $\bar{\nu}_{e}$ fluxes arriving at a detector $D$
can be expressed as $F^{D}_{e}$ and $F^{D}_{\bar{e}}$, respectively:
\begin{equation}\label{eq:fed}
F^{D}_{e}=F^{0}_{e}[P_{h}(P_{l}+P_{2e}(1-2P_{l}))]+
F^{0}_{x}[1-P_{h}(P_{l}+P_{2e}(1-2P_{l}))],
\end{equation}
\begin{equation}\label{eq:fead}
F^{D}_{\bar{e}}=F^{0}_{\bar{e}}[1-(\bar{P}_{l}+\bar{P}_{2e}(1-2\bar{P}_{l}))]
+F^{0}_{x}[\bar{P}_{l}+\bar{P}_{2e}(1-2\bar{P}_{l})]
\end{equation}
for the normal hierarchy, and
\begin{equation}\label{eq:invfed}
F^{D}_{e}=F^{0}_{e}[P_{l}+P_{2e}(1-2P_{l})]
+F^{0}_{x}[1-(P_{l}+P_{2e}(1-2P_{l}))],
\end{equation}
\begin{equation}\label{eq:invfead}
F^{D}_{\bar{e}}=F^{0}_{\bar{e}}[\bar{P}_{h}(1-(\bar{P}_{l}+\bar{P}_{2e}(1-2\bar{P}_{l})))
+F^{0}_{x}[1-\bar{P}_{h}(1-(\bar{P}_{l}+\bar{P}_{2e}(1-2\bar{P}_{l})))]
\end{equation}
for the inverted hierarchy, where
$P_{2e}$ ($\bar{P}_{2e}$) represents the probability that a 
$\nu_{2}$ ($\bar{\nu}_{2}$) arriving at the Earth surface is detected
as a $\nu_{e}$ ($\bar{\nu}_{e}$) at the detector.

The Earth matter density encountered by the
electron neutrinos is given by $V(x)=\sqrt{2}G_{F}N_{e}(x)$, 
where $G_{F}$ is the Fermi constant and
$N_{e}(x)=\rho(x)/(m_{p}+m_{n})$, with $\rho(x)$ the density profile, $m_{p}$ 
the proton mass, and $m_{n}$ the neutron mass. 
The probability $P_{2e}$ 
is given by~\cite{ani:04,ani:05}:  
\begin{equation}\label{eq:p2e}
P_{2e} = \sin^{2}\theta_{12}+\frac{1}{2}\sin^{2}2\theta_{12}
\int_{x_{0}}^{x_{f}} dx [V(x) \sin\phi_{x\rightarrow x_{f}}],
\end{equation}
where $x_{0}$ is the entry point, 
$x_{f}$ is the exit point, and $x$ is the distance into the Earth.
The function $\phi_{a \rightarrow b}$ is defined as 
\begin{equation}\label{eq:phiab}
\phi_{a \rightarrow b}=\int_{a}^{b} \Delta (x) dx,
\end{equation}
where
\begin{equation}\label{eq:delta}
\Delta (x)\equiv\frac{\delta m^{2}_{21}}{2E} 
\sqrt{[\cos2\theta_{12}-\epsilon (x)]^{2}+\sin^{2}2\theta_{12}},
\end{equation}
and the parameter $\epsilon (x)$ is given by
$\epsilon (x) =2EV(x)/\delta m^{2}_{21}$.  
Note that
\begin{equation}
\epsilon(x) \approx 7.575 \times 10^{-8} 
\frac{(E/MeV)(\frac{\rho (x)}{g/cm^{3}})}{\delta m^{2}_{21}/eV^{2}}
\end{equation}
and \begin{equation}
\frac{\delta m^{2}_{21}}{2E} \approx 2.543 \times 10^{3}
\frac{\delta m^{2}_{21}/eV^{2}}{E/MeV} (1/km).
\end{equation}

The density profile of the Earth matter can be approximated 
by a simple step function~\cite{guo:06,freund}:
$\rho (x) \approx 5.0$ $g/cm^{3}$ for $\frac{1}{2} R_{\oplus}<R<R_{\oplus}$ (mantle), 
and $\rho (x) \approx 12.0$ $g/cm^{3}$ for $R< \frac{1}{2} R_{\oplus}$ (core), 
where $R_{\oplus}$ is the Earth radius.
For a specific detector, the incident angle of the
neutrino determines the path length through Earth
and the magnitude of $P_{2e}$.
Because of the
chosen density model, it is necessary to calculate $P_{2e}$
separately for both $0<\psi<\psi_{a}$ (mantle only) and 
$\psi_{a}<\psi<\pi$ (mantle+core+mantle), 
where $\psi$ is the incident angle in general,
and $\psi_{a}$ denotes the incident angle when the
path is tangent to the sphere of radius $R=\frac{1}{2} R_{\oplus}$, at which
radius the density changes.

For $0<\psi<\psi_{a}$, as shown in Fig. 2(a), the probability $P_{2e}$
takes the form
\begin{equation}\label{eq:p2e-1}
P_{2e} = \sin^{2}\theta_{12}+\frac{1}{2}\sin^{2} 2\theta_{12}
[\epsilon (x)(1-\cos(L\Delta (x)))],
\end{equation}
where the total path length $L$ inside the Earth is given by
\begin{equation}\label{L}
L=(d-R_{\oplus})\cos\psi+\sqrt{R^{2}_{\oplus}-
(R_{\oplus}-d)^{2}\sin^{2}\theta_{12}},
\end{equation}
with $d$ the depth of the detector.  

For $\psi_{a}<\psi<\pi$, the path inside the Earth
consists of three segments: $x_{ab}$, $x_{bc}$, and $x_{cd}$,
as shown in Fig. 2(b):

\begin{equation}\label{eq:xab}
x_{ab}=L-(R_{\oplus}-d)\cos\psi-\sqrt{(R_{\oplus}-d)^{2}\cos^{2}\psi-
[(R_{\oplus}-d)^{2}-\frac{R^{2}}{4}]},
\end{equation}
\begin{equation}\label{eq:xbc}
x_{bc}=2\sqrt{(R_{\oplus}-d)^{2}\cos^{2}\psi-
[(R_{\oplus}-d)^{2}-\frac{R^{2}}{4}]},
\end{equation}
\begin{equation}\label{eq:xcd}
x_{cd}=(d-R_{\oplus})\cos\psi-\sqrt{(R_{\oplus}-d)^{2}\cos^{2}\psi-
[(R_{\oplus}-d)^{2}-\frac{R^{2}}{4}]}.
\end{equation}
The probability $P_{2e}$ in this case can be written as
\begin{equation}\label{eq:pec}
P_{2e} = \sin^{2}\theta_{12}+\frac{1}{2}\sin^{2}2\theta_{12} \times I,
\end{equation}
where
\begin{equation}
I=\epsilon^{m}(x)[1-\cos(\Delta^{m} x_{ab})]+
\epsilon^{c}(x)[1-\cos(\Delta^{c} x_{bc})]+
\epsilon^{m}(x)[1-\cos(\Delta^{m} x_{cd})],
\end{equation}
and the superscripts $m$ and $c$ stand for mantle and core, respectively.
It is clear from the oscillatory terms that a longer path length in matter
would give rise to a higher frequency of $P_{2e}$.
Note that for the probability $\bar{P}_{2e}$, the parameter 
$\epsilon(x)=2EV(x)/\delta m^{2}_{21}$ is replaced by $-\epsilon(x)$:
\begin{equation}\label{eq:p2ea}
\bar{P}_{2e}=\sin^{2}\theta_{12}-\frac{1}{2}\sin^{2}2\theta_{12}
\int_{x_{0}}^{x_{f}} dx V(x) \sin(\int_{x}^{x_{f}} dx \frac{\Delta m^{2}_{21}}
{2E}\sqrt{(\cos2\theta_{12}+\epsilon(x))^{2}+\sin^{2}2\theta_{12}}) .
\end{equation}

%%%%%%%%%%%%%%%%%%%%%%%%%%%%%%%%%%%%%%%%%%%%%%%%%%%%%%%%%%%%%%%%%%%%%%%%%%%%%%%
\section{Reexamining the observables with the Earth matter effects included}
%%%%%%%%%%%%%%%%%%%%%%%%%%%%%%%%%%%%%%%%%%%%%%%%%%%%%%%%%%%%%%%%%

For one of the six scenarios $J$,
the expected neutrino flux of flavor $\alpha$ ($\alpha=e,\bar{e}$)
at the surface of Earth, $F^{J}_{\alpha}$,
and at the detector, $F^{D,J}_{\alpha}$, are determined by
the type of mass hierarchy, the values of $P_{h}$,
$P_{l}$ ($\bar{P}_{l}$), and the original
fluxes $F^{0}_{\alpha}$ for a given model.  We list the expressions of $F^{J}_{\alpha}$
and $F^{D,J}_{\alpha}$ in Appendix A.
The effective survival probabilities for $\nu_{e}$ and $\bar{\nu}_{e}$
at the detector, $P^{D,J}$ and $\bar{P}^{D,J}$,
can be read off from 
$F^{D,J}_{e}$ and $F^{D,J}_{\bar{e}}$, respectively:

\begin{eqnarray}\label{eq:effp}
& P^{D,I}=\cos^{2}\theta_{12}+P_{2e}(1-2\cos^{2}\theta_{12}), & \nonumber \\
                  & P^{D,II}=\sin^{2}\theta_{13} \ll 1, & \nonumber \\
& P^{D,III}=P_{2e}, & \nonumber \\
 & P^{D,IV}=P_{2e}, & \nonumber \\
& P^{D,V}=\sin^{2}\theta_{13} \ll 1,  & \nonumber \\
 & P^{D,VI}=\cos^{2}\theta_{12}+P_{2e} (1-2\cos^{2}\theta_{12}), & 
\end{eqnarray}
and
\begin{eqnarray}\label{eq:aeffp}
& \bar{P}^{D,I}=\cos^{2}\theta_{12}-\bar{P}_{2e}(1-2\sin^{2}\theta_{12}), & \nonumber \\
 & \bar{P}^{D,II}=1-\bar{P}_{2e}, & \nonumber \\
& \bar{P}^{D,III}=\sin^{2}\theta_{13} \ll 1, & \nonumber \\
& \bar{P}^{D,IV}=1-\bar{P}_{2e}, & \nonumber \\
& \bar{P}^{D,V}=\cos^{2}\theta_{12}-\bar{P}_{2e}(1-2\sin^{2}\theta_{12}), & \nonumber \\
& P^{D,VI}=\sin^{2}\theta_{13} \ll 1. &
\end{eqnarray}

In principle, the flux variation due to 
the Earth matter effect can be conveniently
illustrated by the ratios $R_{e} \equiv (F^{D}_{e}-F_{e})/F_{e}$ and
$R_{\bar{e}} \equiv (F^{D}_{\bar{e}}-F_{\bar{e}})/F_{\bar{e}}$, 
for $\nu_{e}$ and $\bar{\nu}_{e}$, respectively.
The ratios $R^{J}_{e}$ and $R^{J}_{\bar{e}}$, as given in Appendix B,
suggest the following:

(i) The values of $R^{II}_{e}$, $R^{V}_{e}$, $R^{III}_{\bar{e}}$, 
and $R^{VI}_{\bar{e}}$ vanish, $i.e.$,
including the Earth matter effect would not alter the resultant
$\nu_{e}$ flux for II and V,  and the resultant $\bar{\nu}_{e}$ flux for III and VI.
This can be understood by the fact that the 
crossing probability at the higher resonance approaches 
zero ($P_{h}\sim \sin^{2}\theta_{13} \ll 1$) 
for each of the above scenarios, as indicated by Table 1.  
If the neutrinos travel through the
Earth matter before reaching the detector, Table 1 and
Eq.(46) imply that $P = P^{D,J}=\sin^{2}\theta_{13} \ll 1$ for scenarios II and V.
Likewise, Table 1 and Eq.(47) imply that 
$\bar{P}=\bar{P}^{D,J} = \sin^{2}\theta_{13} \ll 1$ for scenarios III and VI.

(ii) The two ratios, $R^{II}_{\bar{e}}$ and $R^{IV}_{\bar{e}}$, are identical
because of $\bar{P}^{D,II}=\bar{P}^{D,IV}=1-\bar{P}_{2e}$. 
In addition,  one may also conclude that 
$R^{III}_{e} = R^{IV}_{e}$, $R^{I}_{e} = R^{VI}_{e}$,
and  $R^{I}_{\bar{e}} = R^{V}_{\bar{e}}$. 

(iii) The scenarios in each degenerate pair, (II,IV) or (I,VI), 
will not be distinguished from one another by the observation of $\bar{\nu}_{e}$
events alone even if the Earth matter effect is included.  
Thus, constructing the observables that are sensitive to the $\nu_{e}$ events
would be beneficial to the analysis.  Likewise, 
the $\bar{\nu}_{e}$-sensitive observables 
would be needed to lift the degeneracy between III and IV, as well as 
that between I and IV.  

In summary, the values of $R_{e}$ and $R_{\bar{e}}$ may
act as a useful qualitative and quantitative 
reference in analyzing the distinct scenarios. 
For example, scenario II and scenario IV can be distinguished
by examining whether the Earth matter effect is missing 
( scenario II) or observed ( scenario IV) 
in the $\nu_{e}$ flux.  Likewise, scenario III and scenario IV can be separated by
examining whether the Earth matter effect is observed (scenario IV) 
or not (scenario III) in the $\bar{\nu}_{e}$ flux.  Similar qualitative
observations can be applied to separating I and V, or I and VI.

%%%%%%%%%%%%%%%%%%%%%%%%%%%%%%%%%%%%%%%%%%%%%%%%%%%%%%%%%%%%%%

%\subsection{$R^{D}_{1}$}

We now reexamine the expected ratio 
$R^{D}_{1} \equiv N^{D}_{iso}(E<40 MeV)/N^{D}_{iso}(E>40 MeV)$
at a detector when the Earth matter effect is included.
The impact on $R^{D}_{1}$ from the Earth effect can be examined through the plot
$R^{D}_{1}=R^{D}_{1}(\psi)$ in Fig. 3. 
As expected, scenarios III and VI can be easily singled out
by their low
values, $R^{D}_{1} \simeq 2.6$. Note that the values of $R^{D}_{1}$ for III and VI
are nearly independent of the incident angle of neutrinos.  In addition, a jump
of $R^{D}_{1}$ occurs at $\psi \sim \pi/2$ for all the scenarios. 
This is realized by the fact that
the length of neutrino path in mantle begins to
increase significantly when $\psi>\pi/2$.
It is also seen that an even more significant jump occurs at 
$\psi \sim 5\pi/6$, beyond which angle the neutrino flux starts to pass through
the Earth matter that is formed by the combination  
mantle+core+mantle before reaching the detector. 
Note that the values of $R_{1}$ for the four scenarios:
I, II, IV, and V, become almost
indistinguishable when $\psi > 5\pi/6$.

As discussed in Section III, the neutrino events from the $P$-sensitive channels are
needed to remove the degeneracy that the observable $R_{1}$  
fails to resolve. Thus, the $P$-sensitive ratios
$R_{2}\equiv N^{D}_{dir}(E<25 MeV)/N^{D}_{dir}(E>25 MeV)$
as functions of $\psi$ for all the six scenarios are reexamined and
shown in Fig. 4. 

To explain the obvious step-up behavior of 
III and IV, and the step-down behavior of I, II, V, and VI in Fig. 4, one first notes
from Eq.~(\ref{eq:pec}) that $P_{2e}$ increases with $I$.  While the path length
determines the frequency of the oscillatory term, the mean value of
$P_{2e}$ is determined by the sum 
$\epsilon^{m}+\epsilon^{c}+\epsilon^{m}$ for $5\pi/6 < \psi <\pi$,
and by $\epsilon^{m}$ for $0< \psi <5\pi/6$.  
Since the parameter $\epsilon$ is proportional to the density, so that 
$P_{2e}$ increases with the density.  Likewise, Eq.~(\ref{eq:p2ea})
implies that $\bar{P}_{2e}$ decreases when the density increases.
It then follows from Eq.~(\ref{eq:effp}) that $P^{D,I}$ and $P^{D,VI}$ 
decrease when the density increases
$(1-2\cos^{2}\theta_{12}<0)$, while
$P^{D,III}$ and $P^{D,IV}$ increase when the density increases.
On the other hand, Eq.~(\ref{eq:aeffp}) implies that
all $\bar{P}^{D,I}$, $\bar{P}^{D,II}$, $\bar{P}^{D,IV}$, and $\bar{P}^{D,V}$
increase when the density increases.  Given the conditions that 
$P^{D,II}$, $P^{D,V}$, $\bar{P}^{D,III}$, and $\bar{P}^{D,VI}$ are vanishing,
and that the coefficients of the $P$-term dominate over that of 
the $\bar{P}$-term in both
the denominator and the numerator of Eq.~(\ref{r2}), 
the stepping down behavior 
of $R^{D,I}_{2}$, $R^{D,II}_{2}$, $R^{D,V}_{2}$, and $R^{D,VI}_{2}$,
as well as the stepping up of $R^{D,III}_{2}$ and $R^{D,IV}_{2}$ at 
greater $\psi$ become apparent.

%%%%%%%%%%%%%%%%%%%%%%%%%%%%%%%%%%%%%%%%%%%%%%%%%%%%%%%%%%
It should be pointed out, as mentioned earlier, that
the Earth matter effect is manifested by an oscillatory modulation on top
of the neutrino spectra that would have been 
observed without the regeneration effect.  Practically, it would be 
very difficult, if not impossible,
to extract from $R^{D}_{3}$ the information  
that is significantly different from that obtained by $R_{3}$.
%%%%%%%%%%%%%%%%%%%%%%%%%%%%%%%%%%%%%%%%%%%%%%%%%%%%%%%%%%%%%%%%%%%

The ratio of the isotropical events to the directional events,
$R^{D}_{4} \equiv N^{D}_{iso}(13 MeV<E<27 MeV)/N^{D}_{dir}(13 MeV<E<27 MeV)$, 
and the ratio formed by the average
energy of the isotropical to that of the directional events,
$R^{D}_{5} \equiv \langle E \rangle^{D}_{iso}/\langle E \rangle^{D}_{dir}$, 
are shown in Fig. 5 and Fig. 6, respectively.

It is seen that the six scenarios are in principle distinguishable by examining
$R^{D}_{1}$, $R^{D}_{2}$, $R^{D}_{4}$ and $R^{D}_{5}$.  However, as can
be seen from the figures, the distinct scenarios
become less distinguishable at a larger incident angle, $5\pi/6 < \psi <\pi$.
Given the potential theoretical and experimental uncertainties, 
ambiguity may arise in interpreting the proposed observables
if the Earth matter becomes pronounced.

%%%%%%%%%%%%%%%%%%%%%%%%%%%%%%%%%%%%%%
\section{Some results from different supernova models}

We have shown how the original neutrino
fluxes would be modified by the MSW effect in a supernova even if
the details of the density gradient of matter
are unavailable.  The 
input spectral parameters are adopted from the LL model.
As compared to the LL model, the parameters predicted by the ongoing work 
of the Garching group~\cite{Ga,Gb},
labeled as G1 and G2, suggest smaller 
differences in both the flux and in the average energy 
among neutrinos of different flavors:

G1: $\langle E_{\nu_{e}} \rangle = 12 $ MeV,
$\langle E_{\bar{\nu}_{e}} \rangle =15$ MeV, $\langle E_{\nu_{x}} \rangle = 18$ MeV,
$L^{0}_{\nu_{e}}/L^{0}_{\nu_{x}}=L^{0}_{\bar{\nu}_{e}}/L^{0}_{\nu_{x}}=0.8$,

G2:  $\langle E_{\nu_{e}} \rangle = 12 $ MeV,
$\langle E_{\bar{\nu}_{e}} \rangle =15$ MeV, $\langle E_{\nu_{x}} \rangle = 15$ MeV,
$L^{0}_{\nu_{e}}/L^{0}_{\nu_{x}}=L^{0}_{\bar{\nu}_{e}}/L^{0}_{\nu_{x}}=0.5$.

To examine the impact due to the choice of models, 
the expected neutrino spectra for the six possible scenarios
using the parameters of G1 and G2 are shown in Figs. 7 and 8, respectively.  
Note that the spectral shapes predicted by the LL model in 
Fig. 1 exhibit a very distinct nature from that predicted by G1 and G2.
It is seen that 
the peak locations of the isopropical events for III and VI
mark the most significant difference between G1 and G2. 
The peak locations for other scenarios are relatively insensitive 
to the choice of G1 or G2.
As for the directional events, the six scenarios are barely distinguishable
under the G1 model.  In general, the G2 model predicts 
directional spectra that are sharper than 
that predicted by the G1 model.  

The inherent complexity that leads to the entanglement in the spectral shapes, 
the peak heights, 
and the peak locations among scenarios and models may render a thorough, quantitative analysis very 
difficult.  However, as can be seen from Figs. 7 and 8,
the most significant difference among scenarios for a given model
appears in the spectra near the peaks.
Thus, the observed $R^{D}_{1}$ and $R^{D}_{2}$ might be able to
reflect the distinct characters
of each model.  We show the plots of $R^{D}_{1}=R^{D}_{1}(\psi)$ for
G1 and G2 in Fig. 9.  

It is seen that G1
gives rise to distinct values of $R^{D}_{1}(\psi)$ 
ranging from $\sim 11.3$ to $\sim 19.9$.
As for G2, the values of $R^{D}_{1}(\psi)$ are confined within
$36 \sim 37$ for $\psi<\pi/2$, and $28<R^{D}_{1}<37$ for $5\pi/6<\psi<\pi/2$. 
Fig. 9 suggests that if the spectral parameters of G1 are adopted,
the six different scenarios may still cause $R^{D}_{1}$ to vary by roughly a factor
of two.  On the other hand for G2, the ratio $R^{D}_{1}(\psi)$ 
appears to be insensitive to
the uncertainties represented by the distinct scenarios at small incident angle.
As a comparison, one recalls from Fig. 3 that the possible values of
$R^{D}_{1}(\psi)$ for the LL model varies from
$\sim 2.6$ to $\sim 8.8$, which is in a region very different
from that for G1 and G2.  We conclude that the 
observable $R^{D}_{1}$
would reflect the distinct characters of the three different model.  

The ratio $R^{D}_{2}(\psi)$ can be analyzed in a similar way. 
We show the plots of $R^{D}_{2}(\psi)$ for G1 and G2 in Fig. 10.
It is clear that the values of $R^{D}_{2}$ predicted by the three different models 
fall in distinct regions:
$R^{D}_{2}=1.2 \sim 2.5$ for LL, $R^{D}_{2}=2.7 \sim 3.5$ for G1, and 
$R^{D}_{2}=5.3 \sim 6.0$ for G2.
In addition, it should be pointed out that we also perform calculations on 
the other ratios, $R^{D}_{3}$, $R^{D}_{4}$, and $R^{D}_{5}$,
using the input parameters predicted by G1 and G2.
However, unlike $R^{D}_{1}$ and $R^{D}_{2}$, the results show that
the expected values for each of these ratios would all tangled up unless
either a particular model is specified or the uncertainty due to
the density profile can be significantly reduced.

In summary, the choice of models would lead to distinct signatures in 
the values of $R^{D}_{1}(\psi)$ and $R^{D}_{2}(\psi)$.
The degeneracy among the six scenarios for either LL or G1 could be removed.
On the other hand, the very nature of G2
would lead to only very little differences among the scenarios, $i.e.$,
it would be difficult for them to be quantitatively
distinguished.  However, if 
the correct supernova parameters are the ones predicted by the G2 model, 
the analysis based on the G2 model could be relatively free
from the disturbing uncertainties, such as that originating from the density 
profile, the mass hierarchy, and the mixing angle $\theta_{13}$.

%%%%%%%%%%%%%%%%%%%%%%%%%%%%%%%%%%%%%%%%
\section{Summary and conclusions}

While a satisfactory knowledge of the supernova mechanism is still far from
complete, the  
neutrino flux emitted from the core-collapse supernova 
could be a promising tool for probing the unknown neutrino properties.
It is the intention of this work to search for possible hints 
of the unknown neutrino properties using the experimental
observables derived from the supernova neutrino signals.
Given the uncertain density profile and the choice of input spectral models,
the aim is set to exploring the unknown neutrino parameters when the neutrino fluxes 
are modified by the MSW resonant conversion in both the supernova
and in crossing Earth.
Although the MSW mechanism has been well established in the literature,
this analysis shows that the phenomenological outcomes due to the possible 
variation of density profile still provide certain new physical insight
to the nature of the unknown neutrino properties.   

In stead of following the traditional approach based on the 
global density profile $\rho \sim r^{-3}$, 
or the profile with a fixed power in general: $\rho \sim r^{n}$, 
this analysis allows
the density profiles near the locations of resonance to vary
independently as arbitrary power-law functions:
$\rho (r) = c_{l} r^{n_{l}}$ for the lower resonance, and
$\rho (r) = c_{h} r^{n_{h}}$ for the higher resonance.
It is shown that the variation of $n_{k}$ ($k=l$ or $h$) would 
cause different degrees of impact on the survival probability 
for $\nu_{e}$ ($\bar{\nu}_{e}$)
according to whether $n_{k}>-5$ or $n_{k}<-5$.
As far as the survival probabilities are concerned,
there exists six possible scenarios, which are characterized by
possible combinations of the density profile, the mass hierarchy, 
and the mixing angle $\theta_{13}$. 

We propose observable quantities that are derived from the expected
supernova neutrino events at the water Cherenkov detector, and
explore the feasibility of probing the neutrino parameters
with these observables.  As an illustration, the set of spectral
parameters predicted by the Lawrence Livermore group (LL) is adopted.
To investigate how
the choice of supernova models would affect the outcomes,
the analysis based on the input parameters predicted by the Graching
model is also presented as a comparison.

Given the unknown density profile of the supernova and the choice 
of input spectral parameters, we show how the unknown 
mass hierarchy and the mixing angle $\theta_{13}$
would leave observable signatures on the neutrino events. 
The difficulties and limits 
for this type of analysis are also discussed.  
It is shown that the two observables, $R_{1}$ and $R_{2}$, can combine to
act as the most effective pair of observables for probing the neutrino
parameters.
Other observables, $R_{3}$, $R_{4}$, and $R_{5}$, may be considered as the
supplementary references.  
A more realistic analysis incorporating the Earth regeneration effect
is also included.
Figs. 3, 4, 5, and 6 show how each of the observables
varies with the incident angle of neutrinos when
the Earth matter induces additional flavor
conversion prior to the detection of neutrinos.
The general formulation can be further expanded to accommodate
different models of the Earth density profile 
and can be applied to analyzing
the effects  if the neutrinos are registered by other types
of terrestrial detectors. 

%%%%%%%%%%%%%%%%%%%%%%%%%%%%%%%%%%%%%%%%%%%%%%%%
%Note added:
%%%%%%%%%%%%%%%%%

It should be pointed out that certain non-MSW effects
may occur in the complex environment of a 
supernova at different space and time scales.
These effects may be independent of the density profile and
may be totally irrelevant to the MSW conversion.   
If the proper conditions are met, 
recent simulations suggest that effects such as
the neutrino self coupling in dense media
(see, $e.g.$, 
Ref.~\cite{panta,samuel,k:96,fq:06,han:06,duan,duan-1,pastor,raffelt:07}
and the references therein) 
and the neutrino flavor de-polarization associated with
the after shock turbulence 
(see, $e.g.$, Ref.~\cite{benatti:95,loreti:95,fried,flm,choubey:07} and the references therein) 
could occur.  
These effects are expected to impact the efficiency of neutrino flavor transition.
It is suggested  
that both the shock reheating and
the hot bubble epochs could provide the proper conditions for 
two classes of large-scale collective effects: synchronization 
and bipolar flavor transformation, 
which arose if neutrinos themselves form significant background.
The neutrino flavor transformation due to this
neutrino coupling effect shares no common origin 
with that due to the MSW resonance.
In addition, a fluctuating
density in the post-shock region could be created by the turbulent
convective motions behind the forward moving shock and
the reverse shock. Such fluctuations could erase part of the shock wave
imprint on the neutrino spectra even with relatively small amplitudes.
It is suggested that if the amplitude of density 
fluctuations in the turbulence is large, the average neutrino survival 
probability would saturate to 1/2.  The information about the initial state
is then lost, and the final state is depolarized.
This depolarization could leave unique signatures that are distinct from
those left by the shock and other density effects.

As concluded by this work, the variation of density profile, 
which plays a determinant role in the MSW flavor transition, could modify 
the spectra significantly and lead to variation of the proposed observables
by a factor of two or more.
Whether the flavor transition effect arising from one origin
would dominate over that from the others depends on 
the spectral parameters, and on the neutrino intrinsic properties such as
the mass hierarchy and the magnitude of $\theta_{13}$.
In addition, it also depends on the physical conditions 
that the neutrinos would encounter at 
different stages of the supernova environment and in crossing the Earth matter.
Given the undetermined neutrino
intrinsic properties and the existing astrophysical uncertainties, 
it would be intriguing to further investigate, 
under all possible and relevant physical conditions, 
how the non-MSW effects would act as a modification 
to the MSW effect or even dominate over the MSW effect.
The more detailed analyses with reduced assumptions
would certainly help suggest how some of the existing
MSW-based paradigms for the neutrino flavor conversion 
in a supernova should be reconstructed.

%%%%%%%%%%%%%%%%%%%%%%%%%%%%%%%%%%%%%%%%%%%%%%%%%%%%%%%%%%%%%%%%%%

\acknowledgments This work is supported in part by the National 
Science Council of Taiwan.

\appendix 

\section{Expressions for $F^{J}_{\alpha}$ and $F^{D,J}_{\alpha}$}

The neutrino flux of flavor $\alpha$ ($\alpha=e, \bar{e}$)
for the six scenarios at the surface of Earth, $F^{J}_{\alpha}$,
and at the detector, $F^{D,J}_{\alpha}$, are related to the original
fluxes $F^{0}_{\alpha}$ by the following:

\begin{equation}\label{eq:fe1}
F^{I}_{e}=F^{0}_{e}(\cos^{4}\theta_{12}+\sin^{4}\theta_{12})+
F^{0}_{x}[1-(\cos^{4}\theta_{12}+\sin^{4}\theta_{12})],
\end{equation}
\begin{equation}\label{fed1}
F^{D,I}_{e}=F^{0}_{e}[\cos^{2}\theta_{12}+P_{2e}(1-2\cos^{2}\theta_{12})]+
F^{0}_{x}[1-(\cos^{2}\theta_{12}+P_{2e}(1-2\cos^{2}_{12}))],
\end{equation}
\begin{equation}\label{eq:fea1}
F^{I}_{\bar{e}}=F^{0}_{\bar{e}}(\cos^{4}\theta_{12}+\sin^{4}\theta_{12})
+F^{0}_{x}[1-(\cos^{4}\theta_{12}+\sin^{4}\theta_{12})],
\end{equation}
\begin{equation}\label{eq:fead1}
F^{D,I}_{\bar{e}}=F^{0}_{\bar{e}}[1-(\sin^{2}\theta_{12}+\bar{P}_{2e}
(1-2\sin^{2}\theta_{12}))]
+F^{0}_{x}[\sin^{2}\theta_{12}+\bar{P}_{2e}(1-2\sin^{2}\theta_{12})],
\end{equation}
\begin{equation}
F^{II}_{e}=F^{D,II}_{e} = F^{0}_{x},
\end{equation}
\begin{equation}\label{eq:fed2}
F^{II}_{\bar{e}}=F^{0}_{\bar{e}}(\cos^{2}\theta_{12})+
F^{0}_{x}(\sin^{2}\theta_{12}),
\end{equation}
\begin{equation}\label{eq:fead2}
F^{D,II}_{\bar{e}}=F^{0}_{\bar{e}}(1-\bar{P}_{2e})+F^{0}_{x} \bar{P}_{2e},
\end{equation}
\begin{equation}\label{eq:fe2}
F^{III}_{e}=F^{0}_{e} \sin^{2}\theta_{12}+F^{0}_{x} \cos^{2}\theta_{12},
\end{equation}
\begin{equation}\label{eq:fed3}
F^{D,III}_{e}=F^{0}_{e} P_{2e}+F^{0}_{x} (1-P_{2e}),
\end{equation}
\begin{equation}\label{eq:fea3}
F^{III}_{\bar{e}}=F^{D,III}_{\bar{e}}=F^{0}_{x},
\end{equation}
\begin{equation}\label{eq:fe4}
F^{IV}_{e}=F^{0}_{e} \sin^{2}{\theta_{12}}+F^{0}_{x} \cos^{2}\theta_{12},
\end{equation}
\begin{equation}\label{eq:fed4}
F^{D,IV}_{e}=F^{0}_{e}P_{2e}+F^{0}_{x}(1-P_{2e}),
\end{equation}
\begin{equation}\label{eq:fea4}
F^{IV}_{\bar{e}}=F^{0}_{\bar{e}} \cos^{2}\theta_{12}+F^{0}_{x}\sin^{2}\theta_{12},
\end{equation}
\begin{equation}\label{eq:fead4}
F^{D,IV}_{\bar{e}}=F^{0}_{\bar{e}} (1-\bar{P}_{2e})+F^{0}_{x} \bar{P}_{2e},
\end{equation}
\begin{equation}
F^{V}_{e}=F^{D,V}_{e}=F^{0}_{x},
\end{equation}
\begin{equation}
F^{V}_{\bar{e}}=F^{0}_{\bar{e}}(\cos^{4}\theta_{12}+\sin^{4}\theta_{12})
+F^{0}_{x}[1-(\cos^{4}\theta_{12}+\sin^{4}\theta_{12})],
\end{equation}
\begin{equation}
F^{D,V}_{\bar{e}}=F^{0}_{\bar{e}}[\cos^{2}\theta_{12}-\bar{P}_{2e}
(1-2\sin^{2}\theta_{12})]+F^{0}_{x}[\sin^{2}\theta_{12}+
\bar{P}_{2e}(1-2\sin^{2}\theta_{12})],
\end{equation}
\begin{equation}
F^{VI}_{e}=F^{0}_{e}(\cos^{4}\theta_{12}+\sin^{4}\theta_{12})
+F^{0}_{x}[1-(\cos^{4}\theta_{12}+\sin^{4}\theta_{12})],
\end{equation}
\begin{equation}
F^{D,VI}_{e}=F^{0}_{e}[\cos^{2}\theta_{12}+P_{2e}(1-2\cos^{2}\theta_{12})]
+F^{0}_{x}[\sin^{2}\theta_{12}-P_{2e}(1-2\cos^{2}\theta_{12})],
\end{equation}
\begin{equation}
F^{VI}_{\bar{e}}=F^{D,VI}_{\bar{e}}=F^{0}_{x}.
\end{equation}

\section{Flux variation due to Earth matter effect}

Expressions for $R^{J}_{e}$ and $R^{J}_{\bar{e}}$ are given as
the following:

\begin{equation}\label{eq:re1}
R^{I}_{e} =\frac{\sin^{2}\theta_{12}- \cos^{2}\theta_{12}}
            {1+\sin^{4}\theta_{12}+\cos^{4}\theta_{12}}(P_{2e}-\sin^{2}\theta_{12}),
\end{equation}
\begin{equation}\label{eq:rea1}
R^{I}_{\bar{e}} =\frac{3(\sin^{2}\theta_{12}- \cos^{2}\theta_{12})}
            {3(\sin^{4}\theta_{12}+\cos^{4}\theta_{12})+5}
            (\bar{P}_{2e}-\sin^{2}\theta_{12}),
\end{equation}
\begin{equation}
R^{II}_{e} =0,
\end{equation}
\begin{equation}\label{eq:rea2}
R^{II}_{\bar{e}} =\frac{-3}{8-3 \sin^{2}\theta_{12}}
            (\bar{P}_{2e}-\sin^{2}\theta_{12}),
\end{equation}
\begin{equation}
R^{III}_{e}=\frac{1}
            {1+\sin^{2}\theta_{12}}
            (P_{2e}-\sin^{2}\theta_{12}), 
\end{equation}
\begin{equation}
R^{III}_{\bar{e}}=0,
\end{equation}
\begin{equation}\label{eq:re4}
R^{IV}_{e}=\frac{1}
            {1+\sin^{2}\theta_{12}}
            (P_{2e}-\sin^{2}\theta_{12}), 
\end{equation}

\begin{equation}
R^{IV}_{\bar{e}} =\frac{-3}{8-3 \sin^{2}\theta_{12}}
            (\bar{P}_{2e}-\sin^{2}\theta_{12}),
\end{equation}
            \begin{equation}
R^{V}_{e}=0
\end{equation}
\begin{equation}
R^{V}_{\bar{e}}=\frac{3(\sin_{2}\theta_{12}- \cos^{2}\theta_{12})}
            {3(\sin^{4}\theta_{12}+\cos^{4}\theta_{12})+5}
            (\bar{P}_{2e}-\sin^{2}\theta_{12}),
\end{equation}
\begin{equation}
R^{VI}_{e}=\frac{\sin^{2}\theta_{12}- \cos^{2}\theta_{12}}
            {1+\sin^{4}\theta_{12}+\cos^{4}\theta_{12}}(P_{2e}-\sin^{2}\theta_{12}),
\end{equation}
\begin{equation}
R^{VI}_{\bar{e}}=0.
\end{equation}
%%%%%%%%%%%%%%%%%%%%%%%%%%%%%%%%%%%%%%%%%%%%%%%%%%%%%%%%

%%%%%%%%%%%%%%TABLES%%%%%%%%%%%%%%%%

%%%%%table1%%%%%%
 \begin{table}
 \begin{center}
 \begin{tabular}{cccccccccc}  
    $[n_{l}]$ & $P_{l}$ & $\bar{P}_{l}$ & $[n_{h}]$ & $[\theta_{13}]$
    & $P_{h}$ & [mass] & $P$ & $\bar{P}$ & Type  \\ \hline
$<-5$ &  $\cos^{2}\theta_{12}$ &  $\sin^{2} \theta_{12}$ &
$<-5$ & All & 1 & Both  & 
$\cos^{4}\theta_{12}+\sin^{4}\theta_{12}$ & 
 $\cos^{4}\theta_{12}+\sin^{4}\theta_{12}$ & I \\ 
                      $>-5$ & 0 & 0  & $>-5$  & $g>1$  & 0  & Normal  &
                      $\sin^{2}\theta_{13}$  & $\cos^{2}\theta_{12}$  & II  \\ 
  $>-5$ & 0 & 0 & $>-5$ & $g>1$ & 0 &  Inverted & $\sin^{2}\theta_{12}$  & $\sin^{2}\theta_{13}$ & III \\ 
 $>-5$ & 0 & 0 & $>-5$ & $g<1$  & 1  & Both  &  $\sin^{2}\theta_{12}$  
  & $\cos^{2}\theta_{12}$ & IV \\
 $>-5$ & 0  & 0 & $<-5$  & All  &  1  & Both  &  $\sin^{2}\theta_{12}$  &     
  $\cos^{2}\theta_{12}$ & (IV) \\
 $<-5$ & $\cos^{2}\theta_{12}$  & $\sin^{2}\theta_{12}$  & $>-5$  &
 $g>1$  & 0  &  Normal   & $\sin^{2}\theta_{13}$  &  
 $\cos^{4}\theta_{12}+\sin^{4}\theta_{12}$ & V  \\  
 $<-5$ & $\cos^{2}\theta_{12}$ & $\sin^{2}\theta_{12}$ & $>-5$ &
  $g>1$ & 0 &  Inverted  &  $\cos^{4}\theta_{12}+\sin^{4}\theta_{12}$ &
  $\sin^{2}\theta_{13}$  & VI \\
 $<-5$ & $\cos^{2}\theta_{12}$  & $\sin^{2}\theta_{12}$ & 
  $>-5$ &  $g<1$  & 1  &  Both  &  $\cos^{4}\theta_{12}+\sin^{4}\theta_{12}$  &
  $\cos^{4}\theta_{12}+\sin^{4}\theta_{12}$ &  (I)  \\
     \end{tabular}
    \caption{Types of scenarios that are categorized by various combinations of
    $(P,\bar{P})$. Each of the quantity in the square bracket in the first row 
    represents the variable that leads to the probability to its right. 
    The total survival probabilities, $P$ and $\bar{P}$, are determined by
    the specific $n_{l}$, $n_{h}$, $\theta_{13}$, and the mass hierarchy. }
  \end{center}
 \end{table}

%%%%%%%%%%%%FIGURES%%%%%%%%%%%%%%%%%%%%%%%

\begin{figure}
\centerline{\epsfig{figure=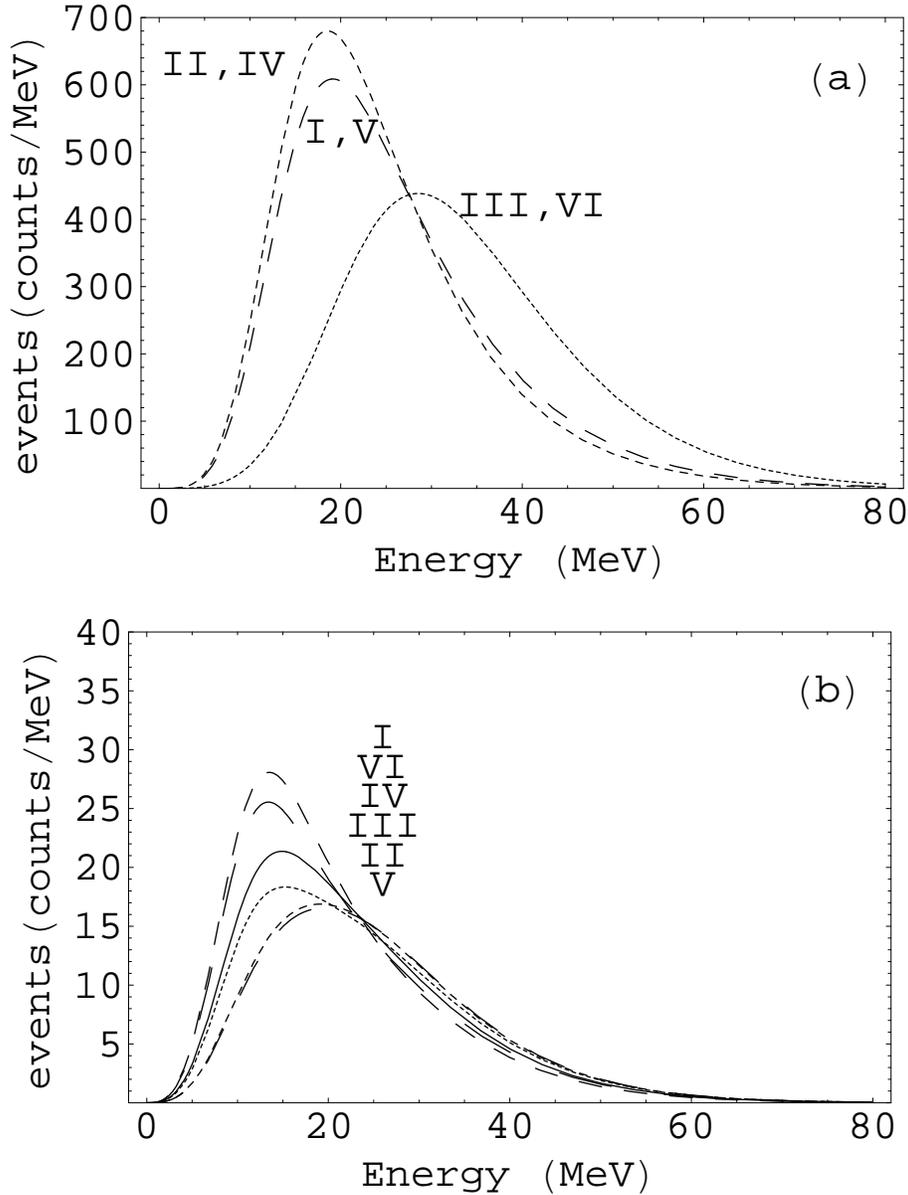,width=5in}}  
\caption{(a) The expected isotropical spectra for the six different scenarios
at a water Cherenkov detector, such as the Super-Kamiokande. 
Note that there are three groups of two-fold degeneracy: (I,V), (II,IV), and (III,VI).
(b) The expected directional spectra for the six scenarios.  
Each curve is labeled according to the height
of its peak.} 
  \label{Figure 1}
    \end{figure} 
    
%%%%%%%fig2%%%%%%%%%%%%%%%%%%%%%%%%%%%%%%%%%%%%%%%%%%

\begin{figure}
\centerline{\epsfig{figure=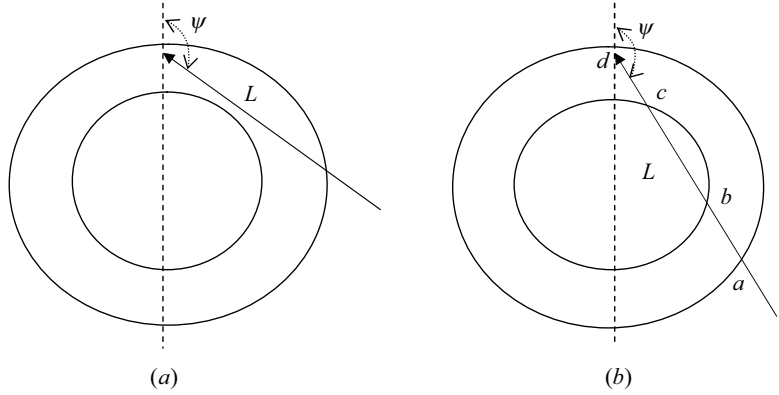,width=6in}}  
\caption{(a) The neutrinos travel through only the mantle of Earth
before reaching the detector.  
(b) The neutrinos travel through the combination of mantle+core+mantle. 
Note that $L$ is the total length
of the path inside Earth and $\psi$ is the incident angle.}
  \label{Figure 2}
    \end{figure} 

%%%%%%%fig3%%%%%%%%%%%%%%%%%%%%%%%%%%%%%%%%%%%%%

\begin{figure}
\centerline{\epsfig{figure=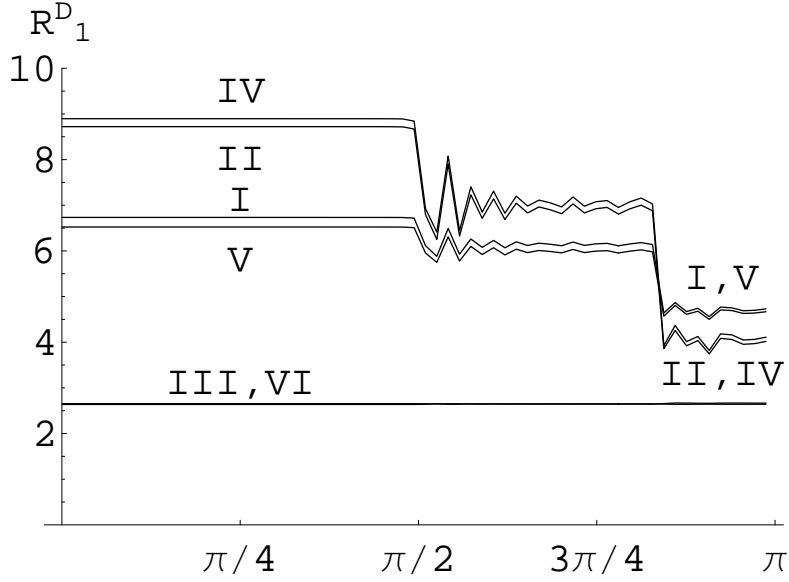,width=5in}}  
\caption{The plot of $R^{D}_{1} \equiv N_{iso}(E<40MeV)/N_{iso}(E>40MeV)$
for all six scenarios as functions of the incident angle $\psi$.
Note that the six scenarios form three nearly degenerate groups.}
  \label{Figure 3}
    \end{figure}

%%%%%%%%%%%%%%%fig4 %%%%%%%%%%%%%%%%%%%%%%%%%%%%%%%%%

\begin{figure}
\centerline{\epsfig{figure=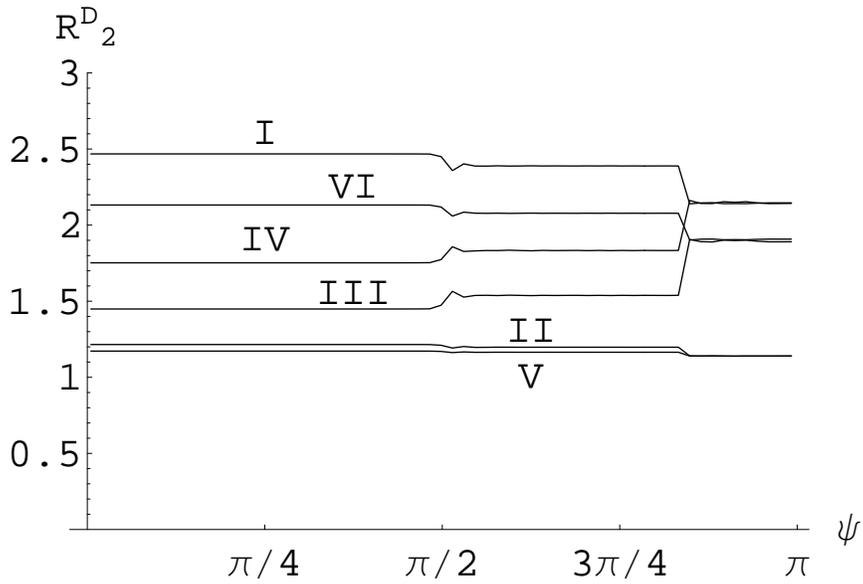,width=5in}}  
\caption{The plot of $R^{D}_{2}\equiv N_{dir}(E<25MeV)/N_{dir}(E>25MeV)$
for all six scenarios as functions of the 
incident angle $\psi$.}
  \label{Figure 4}
    \end{figure}

    %%%%%%%%%%%%%%%%% fig5 %%%%%%%%%%%%%%%%%%%%%%%%%%%%%%%%

\begin{figure}
\centerline{\epsfig{figure=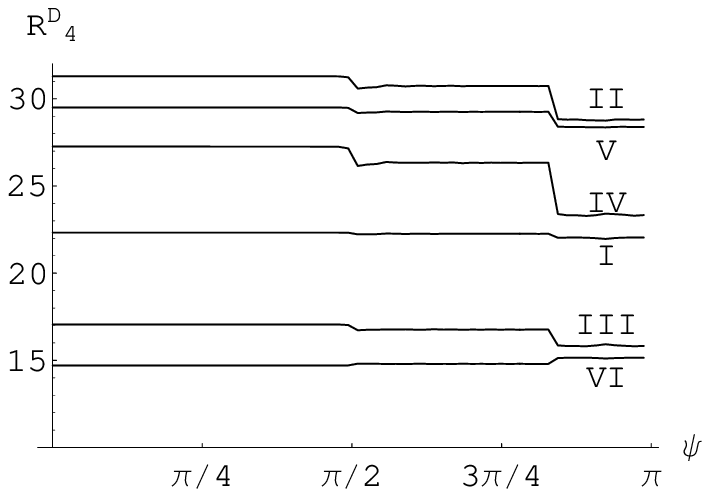,width=5in}}  
\caption{The plot of $R^{D}_{4} \equiv 
N_{iso}(13MeV<E<27 MeV)/N_{dir}(13 MeV<E<27MeV)$ 
for all six scenarios as functions of
the incident angle $\psi$.}
  \label{Figure 5}
    \end{figure} 
    
%%%%%%%%%%%%%%%%%%%%% fig6 %%%%%%%%%%%%%%%%%

\begin{figure}
\centerline{\epsfig{figure=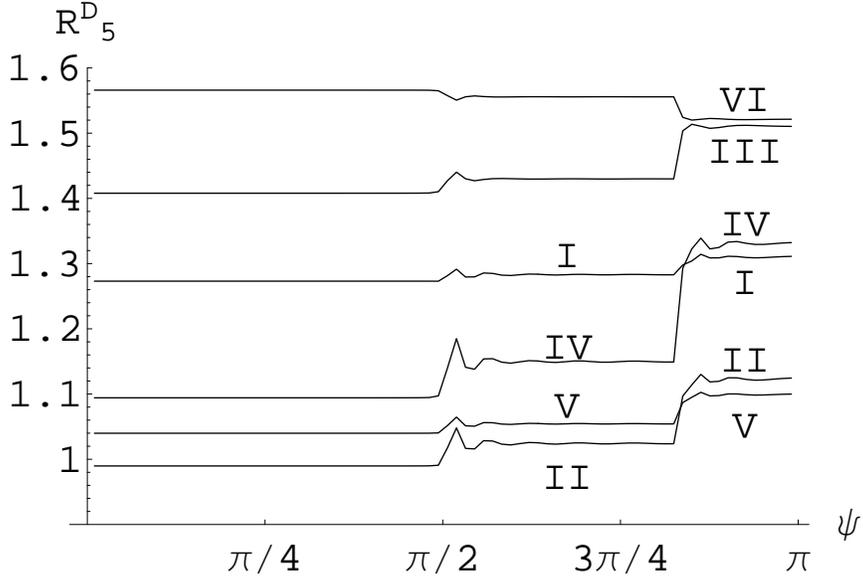,width=5in}}  
\caption{The plot of $R^{D}_{5} \equiv \langle E \rangle_{iso}/
\langle E \rangle_{dir}$ for all six scenarios
as functions of the incident angle $\psi$.}
  \label{Figure 6}
    \end{figure} 
    
%%%%%%%%%%%%%%%%%%%%%%% fig7 %%%%%%%%%%%%%%%%%%%%%%%   
    
\begin{figure}
\centerline{\epsfig{figure=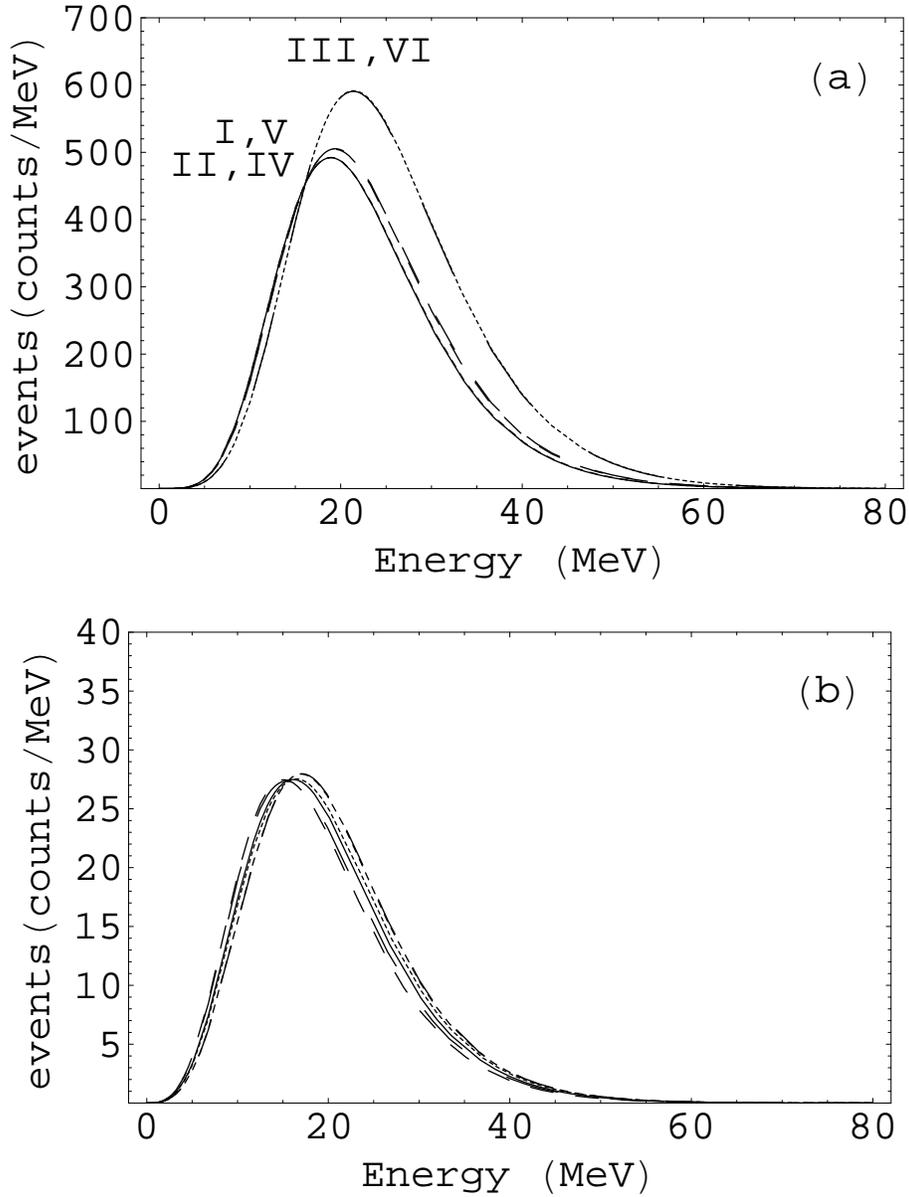,width=5in}}  
\caption{The expected isotropical (a) and directional (b) spectra 
for the six different scenarios predicted by
the G1 model at a water Cherenkov detector.  
There are three groups of two-fold degeneracy in (a): (I,V), (II,IV), and (III,VI).
Note that the curves in (b) are barely distinguishable and are not labeled.} 
  \label{Figure 7}
    \end{figure} 

    %%%%%%%%%%%%%%%%%%%%%%%%%%% fig8 %%%%%%%%%%%%%%%%%%%%
    
\begin{figure}
\centerline{\epsfig{figure=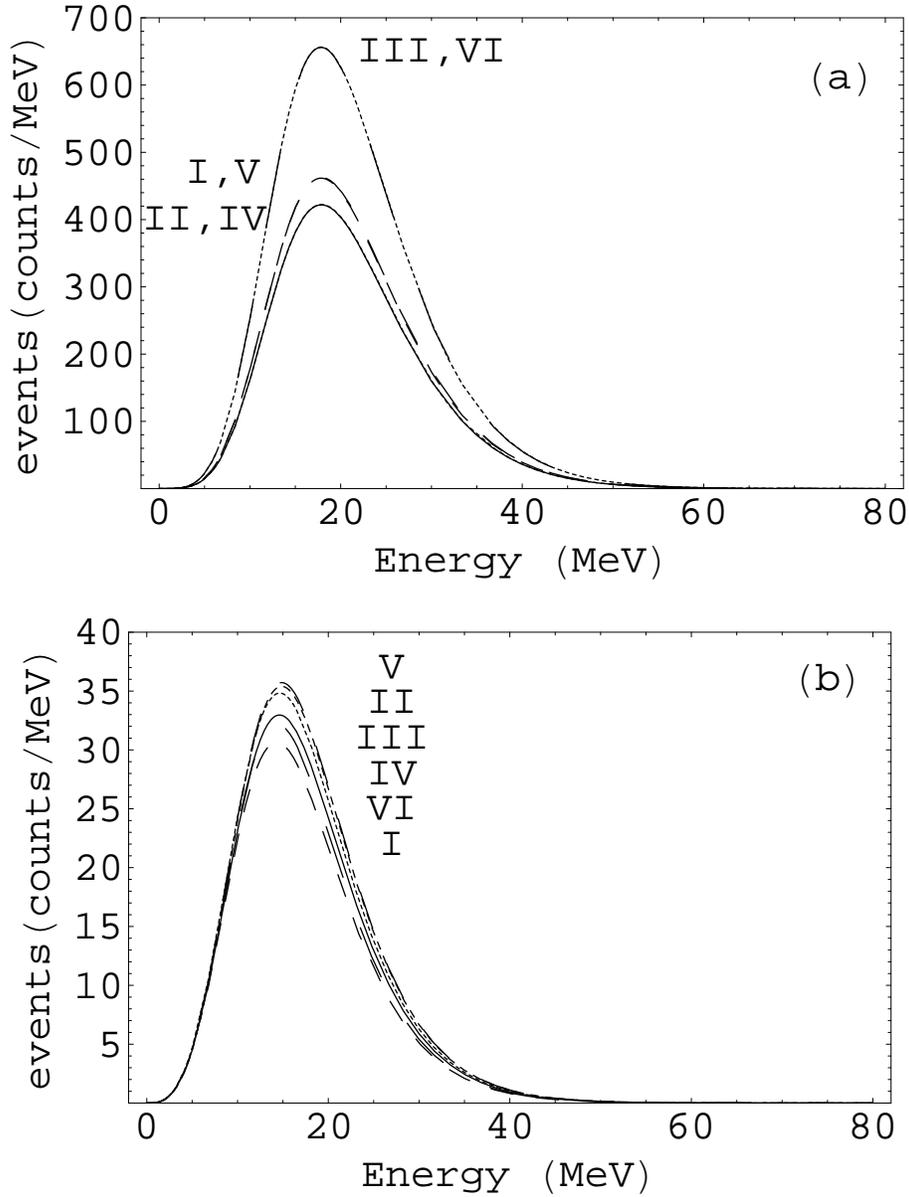,width=5in}}  
\caption{The expected isotropical (a) and directional (b) spectra 
for the six different scenarios predicted by
the G2 model at a water Cherenkov detector.
There are three groups of two-fold degeneracy in (a): (I,V), (II,IV), and (III,VI).  
Each curve in (b) is labeled according to the height
of its peak.} 
  \label{Figure 8}
    \end{figure} 

%%%%%%%%%%%%%%%%%%%%% fig9 %%%%%%%%%%%%%%%%%%%%%%%%%%%%

\begin{figure}
\centerline{\epsfig{figure=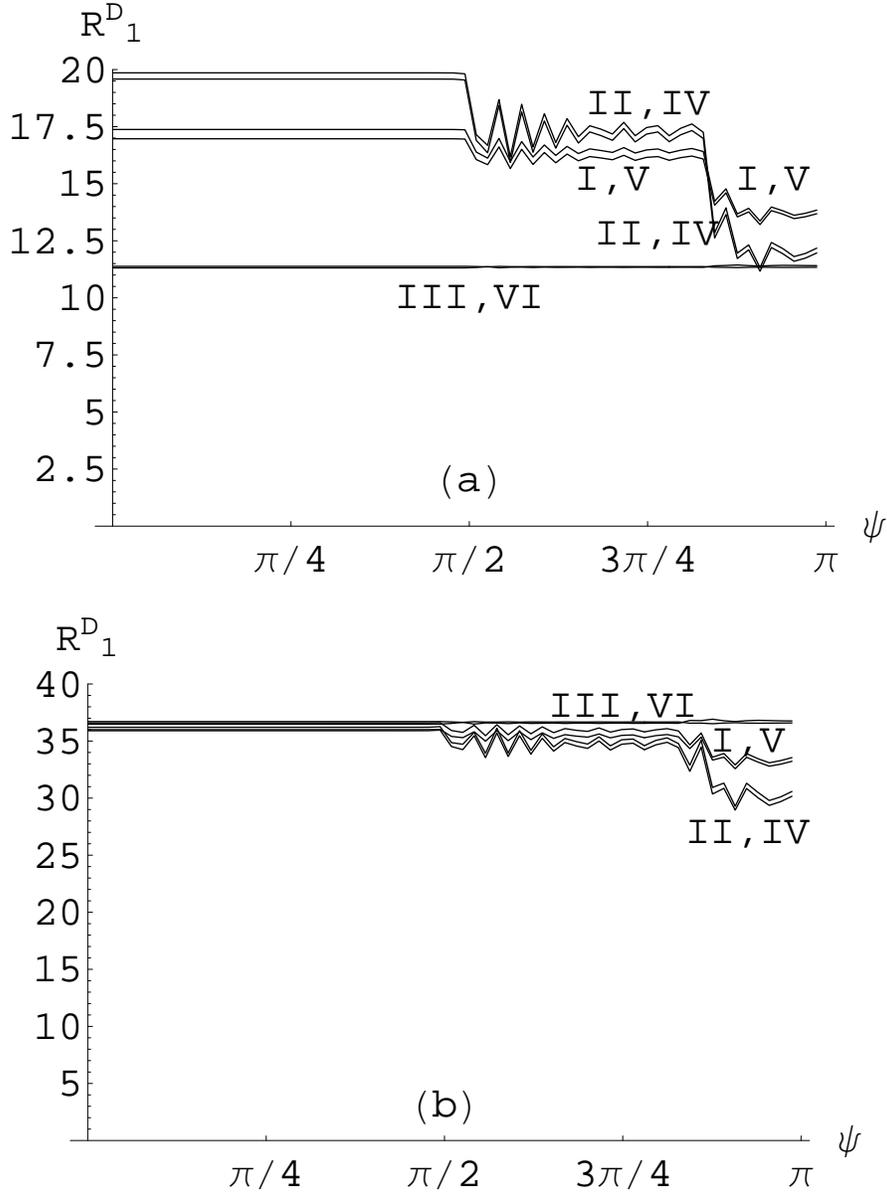,width=5in}}  
\caption{The ratio $R^{D}_{1} \equiv N_{iso}(E<40 MeV)/N_{iso}(E>40 MeV)$
for all the six scenarios
as functions of the incident angle $\psi$ in the G1 model (a), and in the G2 model (b).
Note that $11.3<R^{D}_{1}<19.9$ in (a),
and that the six scenarios are almost indistinguishable
for $\psi< \pi/2$ in (b), $36<R^{D}_{1}<37$.}
  \label{Figure 9}
    \end{figure} 

    %%%%%%%%%%%fig 10 %%%%%%%%%%%%%%%%%%%%%%%%%
    
\begin{figure}
\centerline{\epsfig{figure=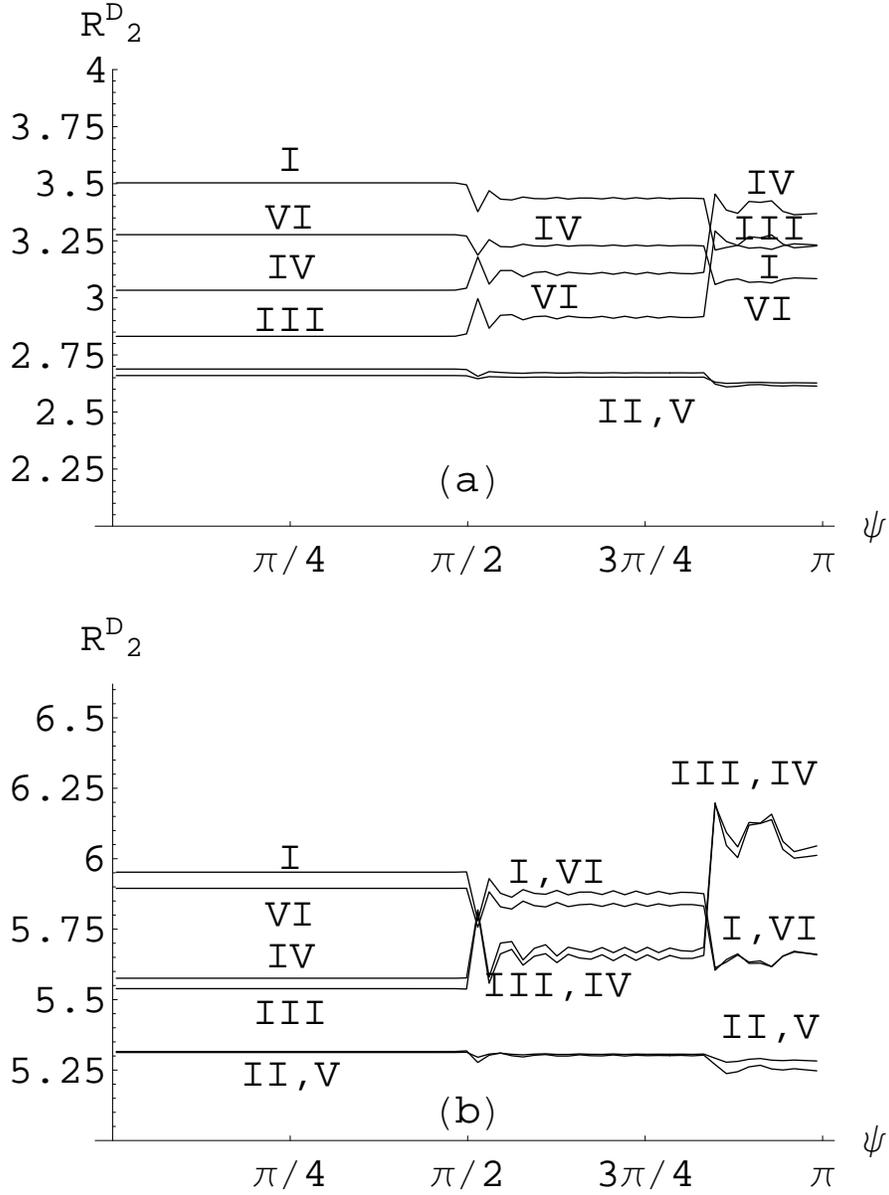,width=5in}}  
\caption{The ratio $R^{D}_{2} \equiv N_{dir}(E<25MeV)/N_{dir}(E>25MeV)$
for all the six scenarios
as functions of the incident angle $\psi$ in the G1 model (a), and in the G2 model (b).
Note that $2.65<R^{D}_{1}<3.50$ in (a),
and that $R^{D}_{1}>5.3$ in (b).} 
  \label{Figure 10}
    \end{figure}

\end{document}